\documentclass{vgtc}                          % final (conference style)
\usepackage{times}                     % we use Times as the main font
\usepackage{wrapfig}
\usepackage{graphicx}
\usepackage[final]{changes}

\usepackage{tabu}                      % only used for the table example
\usepackage{booktabs}                  % only used for the table example
\usepackage{lipsum}                    % used to generate placeholder text
\usepackage{mwe}                       % used to generate placeholder figures

\usepackage{mathptmx}                  % use matching math font

\onlineid{0}

\vgtccategory{Research}

\vgtcinsertpkg

\title{Science-Fiction Transformation as a Design Lens\\for Visualization Transformation}

\author{Abdulhaq Adetunji Salako\thanks{e-mail: abdulhaq.salako@uni-rostock.de} %
\and Christian Tominski\thanks{e-mail: christian.tominski@uni-rostock.de}}
\affiliation{\scriptsize Institute of Visual and Analytic Computing\\University of Rostock\\Rostock, Germany}

\teaser{
  \centering
  \includegraphics[width=\textwidth]{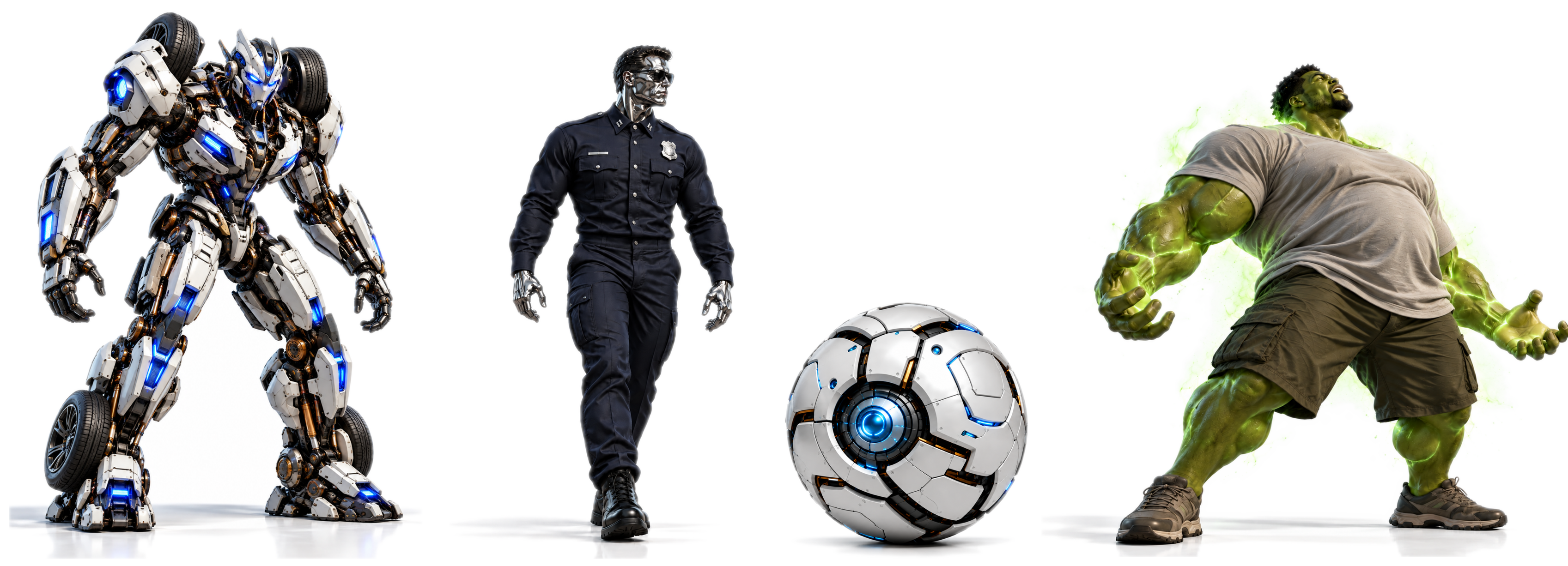}
  \caption{Science-fiction transformations illustrating task adaptation through part reconfiguration, continuous deformation, masked transformation, and bounded amplification.}
  \label{fig:teaser}
}

\abstract{
Science fiction often features characters or objects that undergo radical changes in form while still being recognizable as the same entity. For example, a vehicle might unfold into a robot, or one body might flow or shape-shift into another. We use such transformations as a reflective lens to study similar shape-shifting in visualization, where one visual representation transforms into a completely different one. For example, a node-link diagram might transform into a parallel coordinates plot as a user's task changes from understanding relationships to identifying correlations.
Based on four instances of science-fiction transformation, our paper highlights two key questions: First, why is a transformation into a target form needed, or what is the purpose? Second, what remains recognizable across the transformation from source to target, or what is the correspondence?
Through four hypothetical transitions between visual representations of a multifaceted dataset, we demonstrate that purpose is also essential to visualization transformations. Furthermore, we show that the question of correspondence results in different levels of complexity in terms of designing and understanding transitions. 
Depending on whether correspondence is 1:1, 1:N, or N:1, visualization transformations may be direct and easy to understand, or they may require more elaborate transition constructs, implying increased cognitive effort for interpretation. 
Particularly, cases of $1:N$ and $N:1$ correspondence pose challenges for designing visualization transformations.
}


\nocopyrightspace

\begin{document}

%% The ``\maketitle'' command must be the first command after the
%% ``\begin{document}'' command. It prepares and prints the title block.

%% the only exception to this rule is the \firstsection command

\firstsection{Introduction}
\maketitle

Exploring complex data often requires switching between multiple visual representations, such as node-link diagrams, parallel coordinates, and matrices, each emphasizing different aspects of the same underlying data~\cite{Tominski21FlexibleVA}. Analysts frequently move between views to support comparison, interpretation, and discovery. Yet, each move incurs a context-switching cost, that is, the effort required to leave the context of one view and reorient within another~\cite{Baldonado2000GuidelinesFU}. For example, the same data item may appear as a dot in one representation, a polyline in another, and a more complex shape in yet another. When the view changes, correspondence may not be immediately visible, and must be re-established. This increases the mental effort required to connect representations as data complexity and the number of view types grow.

Visualization research has proposed several ways to reduce this cost. Coordination techniques such as brushing and linking can indicate correspondence between multiple views~\cite{Chen04CompoundBrushing, Steinberger11VisualLinks}. Animation is also often used to convey change within a visualization~\cite{Fisher10Animation}, and it can also support transitions between related visualizations of the same data~\cite{Heer07Animation, Schulz2015Presetbased}. Here, we are interested in animated transformations that change one visualization into a completely different one, for example, a node-link diagram into a parallel coordinates plot~\cite{Salako25AnimTransNLPC}. Such transformations have to be augmented through visible intermediate states, so that the transition itself becomes part of the visual explanation~\cite{Ruchikachorn15Learning}.

Rather than treating such transitions solely as problems of animation, we approach them as design problems concerned with continuity across changing visual forms. A transition may be perceptually smooth while still failing to make the relationship between source and target representations intelligible. This extends attention beyond motion quality to the readability of transformation across marks, layouts, and meaning~\cite{Fisher10Animation, Ruchikachorn15Learning, Tominski21FlexibleVA}. To examine this challenge, we turn to science fiction as a domain in which transformations of form are made visible and interpretable. Science fiction often depicts vehicles, machines, and bodies that change form while remaining recognizable as the same entities~\cite{Troiano16SCIFI}. We do not use science-fiction transformations as literal models for visualization transformations, but as a reflective resource for thinking about how transformations can remain readable.

We develop this argument in four steps. \Cref{sec:related} positions our work in relation to animated visualization transitions and science-fiction-inspired design. \Cref{sec:lens} introduces the aspects used for our reflective analysis, and \Cref{sec:revealDTV} applies these aspects to selected visualization transformations. Finally, \Cref{sec:discussions} discusses the resulting design challenges.

\section{Related Work}
\label{sec:related}

Our work is related to two areas of research: methods for transitioning between visualizations and the use of fiction as a resource for reasoning about visualization transformations.

\subsection{Transition Mechanisms}

Prior work has examined transitions between charts~\cite{Heer07Animation,Fisher10Animation}, multivariate plots~\cite{Yuan09Scatter,Ruchikachorn15Learning}, geographic representations~\cite{Reilly04MapMorphing,Yang2020TiltMI}, and graph visualizations~\cite{Henry07NodeTrix,Schulz2015Presetbased}. Smooth display transitions can also be useful for augmenting
%flexible visual analytics, motivating further work on how transitions can be designed for 
multifaceted visualizations, including spatial (S) and temporal (T) aspects as well as multivariate attributes (A) and relationships (R) among data entities~\cite{Tominski21FlexibleVA}. A smaller body of work has already considered visualization transformations between different data facets~\cite{Windhager20ManyViews, Salako25AnimTransNLPC}. Prior research has articulated design principles and design spaces for animated transitions~\cite{Tversky02Animation, Heer07Animation, Thompson20Animated}, while animation methods such as staging~\cite{Heer07Animation}, staggering~\cite{Chevalier14Staggering}, and bundling~\cite{Du15TrajectoryBundling} aim to make transitions smooth and traceable. Transition grammars and design-space models further specify how views can change and how transitions can be ranked by cost and sequence~\cite{Kim21Gemini, Kim17GraphScapeSimilarity, Salako25VisSimilarity}. Together, these works provide valuable guidance for designing animated transitions between visualizations. Building on this background, we shift the discussion from transition mechanisms to the interpretation of transformation across heterogeneous visual representations.

\subsection{Fiction as a Design Lens}

Science fiction has been used as a resource for reasoning about design. In research on shape-changing interfaces, analyses of science-fiction transformations inform how devices and surfaces may reconfigure their material form~\cite{Troiano16SCIFI, Sturdee17FictionToShapeChanging}. In visualization, fiction has mainly been used for rhetorical, speculative, and critical purposes. Speculative visualization can provoke interpretation and public discussion~\cite{Kim10SpeculativeVisualization}. Fictional superpowers have been used to characterize how visualizations empower people~\cite{willett2021perception}. Science-fiction scenarios have also been used to motivate 3D mobile visualization~\cite{Besancon21MobileVis3D}. Finally, design fiction can be helpful to construct thought experiments that critique visualization at a conceptual level~\cite{Panda24ThoughtExperimentsInDesignFiction}. 

Each of these examples applies fiction to a different end: material shape change, critique, motivation, or rhetoric. Next, we use science-fiction transformations to reason about visualization transformations as a design problem.

\section{Science-Fiction Transformations}
\label{sec:lens}

We already mentioned science-fiction transformations where characters change form to meet new situations (e.g., vehicles unfold into robots). These transformations are presented not as simple substitutions but as staged processes that the audience can follow. Even when the final form differs substantially from the initial one, the passage between them remains legible. This is precisely the challenge that visualization transformations for multifaceted exploration must overcome: maintaining legibility when one visual representation transforms into a substantially different one.

\subsection{Science-Fiction Transformation Examples}

We build our reflective lens from a small set of science-fiction transformations (see \Cref{fig:teaser}) whose general familiarity makes reasoning about them easy. We include both film and game examples because they make transformations understandable in different ways. With films, we merely observe transformations. In games, transformations are often actively triggered for a specific task, which is closer to how analysts invoke visualization transformations.

% Example: https://youtu.be/rtu62Gklgso?si=PDu6Fo_mOG4NQO9V&t=59
\textbf{Transformers}~\cite{wiki:Transformers} -- The Transformers movies involve cars that transform into large humanoid robots. While the car form is suitable for peaceful driving, the robot form is mainly taken for fighting adversaries. During the transformation, a vehicle does not simply vanish and return as a robot but it reorganizes through visible part movements. Doors, wheels, panels, and joints move into new positions, allowing components of the old form to be traced into the new one. The transformation is understandable through a maintained part-to-part correspondence.

% Example: https://youtu.be/aH94xJtqj5A?si=3O-oEIvxZRfaDtxG&t=123
\textbf{Terminator 2}~\cite{wiki:The_Terminator} -- In the Terminator 2 movie, the T-1000's liquid-metal body reshapes through continuous \replaced{deformation}{material and motion} rather than movement of separable parts. \deleted{It is not broken or replaced;} It stretches, deforms, and reforms as one substance. Recognizability rests on material continuity rather than traceable components. The T-1000 adopts a liquid form when its solid humanoid form becomes a hindrance, allowing it to overcome obstacles such as narrow openings. Its body can also deform and reform to recover from bullet impacts. 

% Example: https://www.youtube.com/live/CzHu89Ly45Q?si=nTKhd43ecT_z_tkM&t=1433
\textbf{Metroid's Morph Ball}~\cite{wiki:Metroid} -- The Metroid game allows its protagonist Samus to transform from human form into a ball form. Samus adopts the ball form because it enables movement through spaces her ordinary form cannot enter. Here, the transformation is triggered by the player for a specific task and abandoned when that task is complete. To transform, Samus crouches down and huddles into a ball, but the actual transformation is hidden from the viewer by an energetic cloud. In this case, that fact that Samus and the Morph ball are the same entity can only be inferred from spatio-temporal correspondence: Where Samus was is now the ball.

% Example: https://youtu.be/foa9Gw_KHZw?si=PmC5_xUtKUqQ17LT&t=237
\textbf{The Hulk}~\cite{wiki:Hulk} -- The body of the Hulk changes to meet a demand for greater force and endurance. The transformation involves stretching, expansion, and muscle increase, but overall remains within the logic of the original human body. While the Hulk gains mass and strength during the transformation, it does not become a machine or a fluid. As the form adapts only within the limits of what it can plausibly become (i.e., high correspondence between source and target), the transformation is easy to follow and understand.

\subsection{Relevant Design Aspects}

Looking at these examples, two recurring questions become apparent. First, why does the form change (i.e., purpose)? Second, what stays recognizable through the change (i.e., correspondence)? 

\begin{itemize}
    \item \textbf{Purpose: Why does the form change?} A new form should have a reason to appear. Like the Morph Ball enables passage, a visualization should transition into another representation because a new analytical task calls for it, not merely because change is possible. Purpose is also bounded. Just as the Hulk changes within the logic of the body, a visualization transformation should change within the logic of the data, its visual encoding, and the analytical goal. A transition can only serve the purpose when the new form remains connected to what the data can support and to what the task requires. Conversely, it remains arbitrary when the transformation has no clear analytical or representational reason.

    \item \textbf{Correspondence: What stays recognizable throughout the change?} In science-fiction transformations, recognizability may rest on traceable parts, as in Transformers, or on continuous material and motion, as in the T-1000. In visualization, this question becomes more demanding because the underlying semantic structures and visual encodings may change. For example, when a dot in a node-link diagram is transformed into a polyline in a parallel coordinates plot, there is no persistent mark to follow, there is not even a common frame of reference. So, the correspondence between old and new visual forms needs to be purposefully designed to be perceivable and understandable.
\end{itemize}

These aspects are not independent of how transformations are shown. In fact, a transformation is recognizable only if enough of the process is visible. Therefore, visibility is the enabling condition through which purpose and correspondence become readable. This thinking resonates with prior analyses of transformation in science fiction~\cite{Troiano16SCIFI}.

Together, purpose, correspondence, and visibility provide the reflective lens through which we examine visualization transformations in the next section.

% Station from Bill and Ted’s Bogus Journey https://youtu.be/pLqoZtmyxxk?t=67
% The Fly (1986) 
% Big Hero 6 https://www.youtube.com/watch?v=SOqFO2NexSk
% Transformers in KSI https://youtu.be/grcPt3jw1rI?t=239

\section{Understanding Visualization Transformations}
\label{sec:revealDTV}

To see how the aforementioned aspects influence the design of visualization transformations, we consider the scenario of exploring multifaceted data. Assume the data represents a collaboration network consisting of authors, their co-authorship relations (R), associated attributes (A), geo-spatial locations (S), and publication timelines (T). Designing a visualization that shows all data facets in full detail is usually difficult, if not impossible~\cite{Tominski21FlexibleVA}. As a thought experiment~\cite{Panda24ThoughtExperimentsInDesignFiction}, we will next outline four hypothetical visualization transformations that may play a role when analysts explore multifaceted data using multiple views. 

Much like in science fiction, we envision \emph{transformable visualizations} that can shape-change depending on analytical focus. Initially, we start with a node-link diagram that visualizes the relational structure (R) with authors being represented as dots and co-authorships as links connecting pairs of dots.
The node-link diagram is our common source representation, which will be transformed through four hypothetical transformations into different visual representations that shift the analytical focus regarding each data facet. In other words, each transformation changes the node-link representation into a different visualization with a different analytical purpose.

\subsection{Visualization Transformation Examples}

Next, we look at the four example transformations in detail. We examine why the target representations are appropriate for certain purposes, and what correspondences must remain recognizable as the transformations unfold.

\setlength{\intextsep}{5pt}
\setlength{\columnsep}{5pt}

\begin{wrapfigure}{r}{0.27\columnwidth}
  \centering
  \includegraphics[width=0.26\columnwidth]{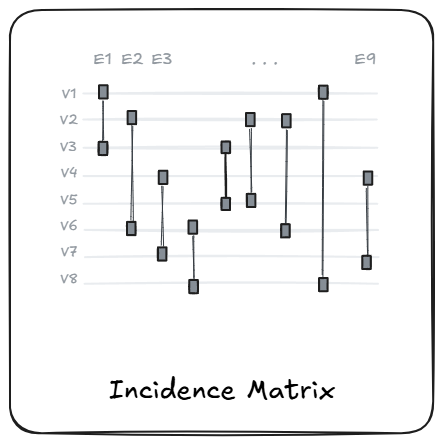}
\end{wrapfigure}
\textbf{Node-link to incidence matrix.} As link density increases, a node-link view becomes cluttered and difficult to read. Transforming it to an incidence matrix (aka BioFabric~\cite{Longabaugh12BioFabric}) replaces the force-directed layout with an ordered arrangement, in which nodes are represented as horizontal rows, and edges are represented by incidence marks instead of crossing links.
\emph{Purpose:} This transformation is needed by a shift from tracing dense links in a force-directed layout to interpreting relationships through structured alignment.
\emph{Visibility:} Nodes, originally drawn as a dots, are visibly reshaped into horizontal rows that are positioned according to an assigned matrix order. Links are then transformed into vertical incidence marks that connect the rows of their incident nodes. This reorganization reflects the part reconfiguration seen in the Transformers movies.
\emph{Correspondence:} Each node and its row as well as each link and its incidence mark are in a $1:1$ correspondence. The incidence marks preserve the relational identity of the co-author collaborations by continuing to link the same pair of authors.

\begin{wrapfigure}{r}{0.27\columnwidth}
  \centering
  \includegraphics[width=0.26\columnwidth]{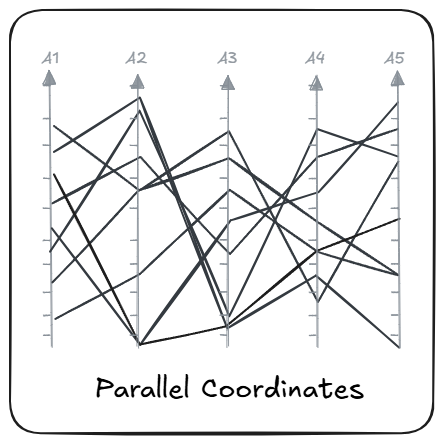}
\end{wrapfigure}
\textbf{Node-link to parallel coordinates.} While a node-link diagram effectively represents relationships among authors, it provides limited support for analyzing their multivariate attributes. Parallel coordinates~\cite{Inselberg09PC} instead represent each author's attributes across a shared set of parallel axes, enabling detailed inspection of trends and correlations.
\emph{Purpose:} The analytical focus changes from relationships (R) among authors to investigations of author attribute values (A) across multiple dimensions.
\emph{Visibility:} To support this change, the visualization transformation unfolds the dots representing authors in the node-link diagram into polylines that bend according to the authors' attribute values~\cite{Salako25AnimTransNLPC}, similar to the continuous deformation in Terminator 2. The transformation further fades in a set of parallel axes as a new frame of reference.
\emph{Correspondence:} During the transformation, each dot representing an author becomes a polyline. The correspondence is therefore $1:1$. Because parallel coordinates provide no graphical encoding for relationships, links are not preserved.

\begin{wrapfigure}{r}{0.27\columnwidth}
  \centering
  \includegraphics[width=0.26\columnwidth]{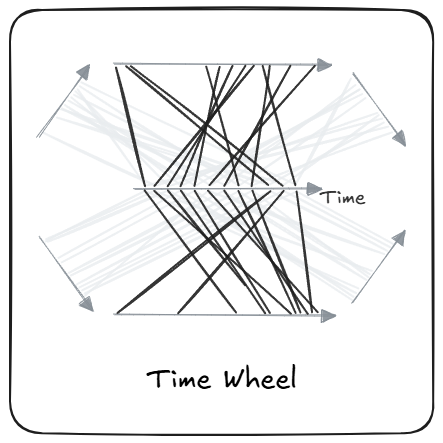}
\end{wrapfigure}
\textbf{Node-link to time wheel.} As our node-link diagram does not reveal how publications are distributed over time and across publication types, we transform into the time wheel~\cite{Tominski04AxesBasedVis}, which shows time (T) and attributes (A). Similar to parallel coordinates, the time wheel is based on axes, but it has a central time axis around which several axes for time-dependent attributes (e.g., counts per publication type) are arranged.
\emph{Purpose:} As such, the time wheel allows us to see temporal publication patterns across publication types, while omitting relationships among authors. 
\emph{Visibility:} The transformation from node-link diagram to time wheel first fades out the links (R is not encoded in the time wheel). For all authors, their dot representation is then split into a set of line segments, one segment per publication type. The line segments are then smoothly arranged such that they connect the central time axis and their corresponding publication-type axis to represent the count per publication type.
\emph{Correspondence:} While each author's visual mark changes from a single dot to multiple line segments, which could be considered a bounded amplification as in The Hulk, the publications remain associated with the same author. So, the transformation yields a $1:N$ correspondence between an author's dot and its publication-type line segments. 

\begin{wrapfigure}{r}{0.27\columnwidth}
  \centering
  \includegraphics[width=0.26\columnwidth]{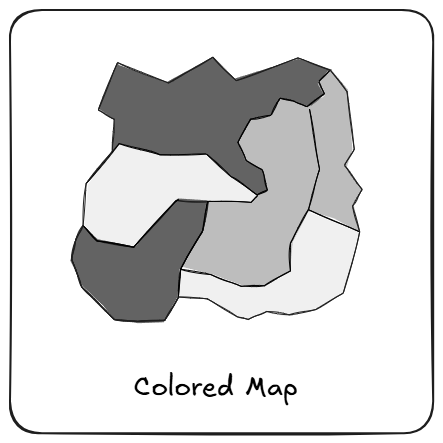}
 \end{wrapfigure}
\textbf{Node-link to colored map.} Finally, we consider the spatial (S) facet of the data and use a colored map to reveal how many authors are located per geographic region. Transforming to a map replaces the force-directed layout of dots with geographic regions, where spatial location rather than network structure determines the visual arrangement.
\emph{Purpose:} Understanding the spatial distribution of authors across geographic regions is now the analytical focus.
\emph{Visibility:} During the envisioned transformation, each author's dot moves from its original node-link position to the center of its associated geographic region. At each center, multiple author dots visibly merge as they accumulate in the same spot. This co-location merging without actual change in form reflects the masked transformation seen in \textit{Metroid}. Finally, each region's merged dot is shape-shifted into the form of the geographic region. 
\emph{Correspondence:} The correspondence in this case is $N:1$, with multiple individual author dots being transformed into one geographic area.

\subsection{What Can We Learn?}

The outlined hypothetical transformations illustrate the role of the \emph{purpose} in reasoning about visualization transformations. In our examples, the transformations redirect analytical attention from the R data facet to the A, S, or T facets, or reinterpret R through a different representation. This suggests that any transformation begins with a change in analytical purpose that determines which data facets should remain relevant, become emphasized, or are no longer represented.

\begin{table}[t]
    \centering
    \caption{Correspondence cardinality as an indicator of expected transformation complexity for designers and users.}
    \label{tab:correspondence}
    \begin{tabular}{lccc}
        \textbf{Correspondence} & $1:1$ & $1:N$ & $N:1$ \\
        \noalign{\vspace{3pt}}\hline\noalign{\vspace{3pt}}
        \textbf{Transition complexity} (Designer) & Direct & \multicolumn{2}{c}{Constructed} \\
        \textbf{Understandability} (User) & Easy & \multicolumn{2}{c}{Difficult} \\
    \end{tabular}
\end{table}

As the analytical purpose changes, visualization transformations have to ensure the \emph{visibility} of the referential structures (i.e., axes) required for the involved data facets. These structures provide the visual context for interpreting the data in the transformed representation. Conversely, referential structures that no longer support the analytical purpose can be omitted (e.g., the links).  
With the referential structures established, the marks representing the data items can be transformed from the source and to the target representation. Various ways of actually performing the transformation are possible. Here, we specifically saw resemblance to science-fiction transformations, but many others could be devised to make changes from source to target representation visible.

The source-to-target \emph{correspondence} may vary across transformations and has an influence on the complexity of the transformation process. \Cref{tab:correspondence} summarizes possible correspondences at the level of data items and their associated visual marks. For $1:1$ correspondence, each data item and its mark in the source correspond to a single item and mark in the target. For visual marks, correspondence also depends on geometric compatibility. When source and target have compatible geometry, the target mark can emerge directly from the source. A dot, for instance can extend into a matrix row that passes through it's position. With incompatible geometries, intermediate steps may be needed. A dot may first need to be split into arc segments before these can be unfolded into a polyline.

As correspondence changes to $1:N$ or $N:1$, additional transition constructs are required. For $1:N$, a single data item represented by one mark in the source must be split into multiple marks in the target. For $N:1$, multiple data items, each with its own mark in the source, must be combined into a single mark in the target. These forms of correspondence make splitting and merging more difficult to communicate and may increase the effort required to design and interpret the transformations. Correspondence cardinality is therefore a helpful indicator for transformation complexity.

\section{Discussion}
\label{sec:discussions}

Next, we discuss the implications of correspondence cardinality for transformation design and interpretation, and reflect on limitations of our science-fiction lens on visualization transformations.

\paragraph{Design and Interpretation}

Considering correspondence cardinality shifts the design emphasis from transforming marks to communicating correspondence between marks. For $1:N$ and $N:1$ cardinality, designing animated transitions requires more than smoothly transforming individual marks. Additional transition constructs are needed to establish one-to-many and many-to-one correspondence between source and target marks. The same challenge extends to interpretation, as viewers must infer these relationships because they are not established by continuous transformation alone. Transformation complexity therefore depends not only on preserving continuity, but also on maintaining understandable correspondence as the mapping changes. Yet, correspondence cardinality is only an indicator of transformation complexity, not a fixed rule. For example, a higher-cardinality mapping may remain easy to understand when split and merge follow a clear visual and semantic structure, whereas a one-to-one mapping may still be perceived as rather complex if source and target marks differ strongly in geometry, encoding, or referential structure.

Higher-cardinality correspondence can be made more understandable by visually grouping the marks participating in the same correspondence or by preserving visible cues that link corresponding source and target marks throughout or after the transition. Such cues allow correspondence to remain observable instead of relying entirely on the viewer's tracking of individual mark movements during the transformation. These observations raise questions about when such visual aids improve understanding, how they can be introduced without adding visual clutter, and how transitions should be evaluated when communicating correspondence becomes as important as animation smoothness and perceptual traceability.

\paragraph{Extending to Transformation Sequences.}

The discussion so far has considered correspondence cardinality within individual visualization transformations. Beyond basic animated transitions, transformable visualizations as we envision them here may involve transformation sequences. In such sequences, the target representation of one transformation becomes the source representation of the next transformation~\cite{Tominski21FlexibleVA}. Correspondence may therefore change across the sequence rather than being established independently for each transformation. For example, a $1:1$ correspondence may later become $1:N$ and subsequently $N:1$ as different representations are introduced. This suggests that the correspondence established in one transformation may influence the design and interpretation of subsequent transformations, making correspondence across transformation sequences a useful consideration.

\paragraph{Limitations.} 

Our selected science-fiction transformations are illustrative rather than exhaustive. They demonstrate how science fiction can serve as a lens for discussing design aspects of visualization transitions, but they represent only a subset of a much larger design space. From a broader set of examples, recurring transformation strategies could be identified and generalized for use in transformable visualizations.

Our hypothetical visualization transformation cover $1:1$, $1:N$, and $N:1$ correspondences, leaving $N:N$ correspondence unexplored. We may find examples of such complex transformations in science fiction (e.g., swarm-based deformations). However, the usefulness of $N:N$ transitions for visualization seems to be low in light of the limits of the human cognitive system.

Finally, our visualization transformations are mostly ad-hoc examples, and their full design details remain elusive. Exploring additional transformation types and transition strategies would certainly broaden our understanding of the design space of visualization transitions.

\section{Conclusion}

We used science-fiction transformation as a reflective lens for reasoning about visualization transformations. This perspective advocates visibility and frames two design questions: Why is a new representation needed and what must remain recognizable as one representation changes into another.

Applying this perspective to four hypothetical visualization transformations showed that purpose motivates the choice of the target representation, while smooth animated transitions make changes visible. Correspondence determines what may change and what must remain recognizable. Correspondence cardinality emerged as a useful indicator of transformation complexity: $1:1$ correspondences can often be realized through the transformation of a single mark, whereas $1:N$ and $N:1$ correspondences require additional transition constructs to communicate correspondence. Although the examples are illustrative, they demonstrate how science fiction can serve as a lens for reasoning about transformable visualizations and suggest opportunities to explore additional transformation scenarios.

% \section*{Supplemental Materials}
% \label{sec:supplemental_materials}

% Refer to the instructions for this section (\cref{sec:supplement_inst}).
% Below is an example you can follow that includes the actual supplemental material for this template:

% All supplemental materials are available on OSF at \url{https://doi.org/10.17605/OSF.IO/2NBSG}, released under a CC BY 4.0 license.
% In particular, they include (1) Excel files containing the data for and analyses for creating \cref{tab:vis_papers} and \cref{fig:vis_papers}, (2) figure images in multiple formats, and (3) a full version of this paper with all appendices.
% Our other code is intellectual property of a corporation---Starbucks Research---and there is no feasible way to share it publicly.

% \section*{Figure Credits}
% \label{sec:figure_credits}

% Refer to the instructions for this section (\cref{sec:figure_credits_inst}).
% Here are the actual figure credits for this template:

% \Cref{fig:teaser} image credit: Scott Miller / Special to the Vancouver Sun, January 22, 2009, page A6.

% \Cref{fig:vis_papers} is a partial recreation of Fig.\ 1 from \cite{Isenberg:2017:VMC}, which is in the public domain.

%% if specified like this the section will be committed in review mode
\acknowledgments{
We thank the Deutsche Forschungsgemeinschaft (DFG) for supporting this research under grant number \href{https://gepris.dfg.de/project/514630063}{514630063}. \Cref{fig:teaser} is an AI-generated illustration produced with a text-to-image tool to depict conceptual examples.}

\bibliographystyle{abbrv-doi-hyperref}

\bibliography{template}
\end{document}